\documentclass[letterpaper,twocolumn,10pt]{article}
\usepackage{usenix2019_v3,epsfig}
\usepackage{epigraph}
\usepackage{makecell} 
\usepackage{multirow}
\usepackage{xspace}
\usepackage{paralist}
\usepackage{libertine}
\usepackage{eurosym}
\usepackage{amssymb}
\usepackage{xcolor}
\usepackage{graphicx}
\usepackage{authblk}

\definecolor[named]{MyRed}{cmyk}{0,0.84,0.8,0.19}
\definecolor[named]{MyGreen}{cmyk}{0.89,0,0.66,0.60}

\usepackage{booktabs} 

\newcommand{\sref}[1]{\S\ref{#1}}
\newcommand{\vheading}[1]{\vspace{0.05in}\noindent\textbf{#1}}
\newcommand{\viheading}[1]{\vspace{0.05in}\noindent\emph{#1}}

\newcommand{\eg}{\textit{e.g.,}\xspace}

\newcommand{\myx}{$\times$\xspace}

\begin{document}

\date{}

\title{\LARGE \bf Analyzing the Impact of GDPR on Storage Systems\thanks{\textsuperscript{*}Aashaka Shah and Vinay Banakar contributed equally.}}
\renewcommand\footnotemark{}

\author[1]{Aashaka Shah\textsuperscript{*}}
\author[3]{Vinay Banakar\textsuperscript{*}}
\author[1]{Supreeth Shastri}
\author[2]{Melissa Wasserman}
\author[1]{Vijay Chidambaram}
\affil[1]{\normalsize \it Computer Science, University of Texas at Austin}
\affil[2]{\normalsize \it School of Law, University of Texas at Austin}
\affil[3]{\normalsize \it Hewlett Packard Enterprise}


\maketitle

\pagestyle{empty}

The recently introduced General Data Protection Regulation (GDPR) is
forcing several companies to make significant changes to their
systems to achieve compliance. Motivated by the finding that 
more than 30\% of GDPR articles are related to storage, we investigate 
the impact of GDPR compliance on storage systems. We illustrate the 
challenges of retrofitting existing systems into compliance by 
modifying Redis to be \emph{GDPR-compliant}. We show that 
despite needing to introduce a small set of new features, a strict 
real-time compliance (\eg logging every user request synchronously) 
lowers Redis' throughput by 20$\times$. Our work reveals how GDPR allows 
compliance to be a spectrum, and what its implications are for 
system designers. We discuss the technical challenges that need to 
be solved before strict compliance can be efficiently achieved.

\section{Introduction}
\label{sec-introduction}

\setlength{\epigraphwidth}{2.1in}
\setlength{\epigraphrule}{0.1pt}
\epigraph{\emph{``In law, nothing is certain but the expense.''}}{Samuel Butler}

Privacy and protection of personal data (or more aptly, the lack
thereof) has become a topic of concern for the modern society. The
gravity of personal data breaches is evident not only in their
frequency ($\sim$1300 in 2017 alone~\cite{data-breaches-2017}) but
also their scale (the Equifax breach~\cite{equifax} compromised the
financial information of $\sim$145 million consumers), and scope (the
Cambridge Analytica scandal~\cite{cambridge-analytica} harvested
personal data to influence the U.K. Brexit referendum and the 2016
U.S. Presidential elections). In response to this alarming trend, the
European Union (EU) adopted a comprehensive privacy regulation called
the General Data Protection Regulation (GDPR)~\cite{gdpr-regulation}.

GDPR defines the privacy of personal data as a fundamental right of
all European people, and accordingly regulates the entire
lifecycle of personal data. Thus, any company dealing with EU people's
personal data is legally bound to comply with GDPR. While essential,
achieving compliance is not trivial: Gartner
estimates~\cite{gartner-prediction} that less than 50\% of the
companies affected by GDPR would likely be compliant by the end of
2018. This challenge is exacerbated for a vast majority of companies
that rely on third-parties for infrastructure services, and hence, do
not have control over the internals of such services. For example, a
company building a service on top of Google cloud storage system would 
not be compliant if that cloud subsystem is violating the GDPR norms. 
In fact, GDPR prevents companies from using any third-party services 
that violate its standards.

Though GDPR governs the behavior of most of the infrastructure and 
operational components of an organization, its impact on the storage 
systems is potent: 31 of the 99 articles that make up GDPR
directly pertain to storage systems. Motivated by this finding, we 
set out to investigate the impact of GDPR on storage systems. In 
particular, we ask the following questions: (i) What features should 
a storage system have to be GDPR-compliant? (ii) How does 
compliance affect the performance of different types of storage 
systems? (iii) What are the technical challenges in achieving strict 
compliance in an efficient manner?

By examining the GDPR articles, we identify a core set of (six) 
features that must be implemented in the storage layer to achieve 
compliance. We hypothesize that despite needing to support a small 
set of new features, storage systems would experience a significant
performance impact. This stems from a key observation: GDPR's goal of 
\emph{data protection by design and by default} sits at odd with 
the traditional system design goals (especially for storage systems) 
of optimizing for performance, cost, and reliability. For example, 
the regulation on identifying and notifying data breaches requires 
that a controller shall keep a record of all the interactions with 
personal data. From a storage system perspective, this turns every 
read operation into a read followed by a write. 

To evaluate our hypothesis, we design and implement the changes 
required to make Redis, a widely used key-value store, \emph{
GDPR-compliant}. This not only illustrates the challenges of 
retrofitting existing systems into GDPR compliance but also quantifies 
the resulting performance overhead. Our benchmarking using YCSB
demonstrates that the GDPR-compliant version experiences a 20$\times$
slowdown compared to the unmodified version. 

We share several insights from our investigation. First, though GDPR
is clear in its high-level goals, it is intentionally vague in its 
technical specifications. This allows GDPR compliance to be a 
continuum and not a fixed target. We define \emph{real-time 
compliance} and \emph{eventual compliance} to describe a system's 
approach to completing GDPR tasks. Our experiments show the 
performance impact of this choice. For example, by storing the 
monitoring logs in a batch (say, once every second) as opposed to 
synchronously, Redis' throughput improves
by 6\myx while exposing it to the risk of losing one second worth of 
logs. Such tradeoffs present design choices for researchers and 
practitioners building GDPR-compliant systems. Second, some GDPR
requirements sit at odds with the design principles and performance
guarantees of storage systems. This could lead to storage systems
offering differing levels of native support for GDPR compliance (with
missing features expected to be handled by other infrastructure or
policy components). Finally, we identify three key research challenges
(namely, efficient deletion, efficient logging, and efficient metadata
indexing) that must be solved to make strict compliance efficient.

\section{Background on GDPR}
\label{sec-gdpr}

GDPR~\cite{gdpr-regulation} is laid out in 99 \emph{articles} that 
describe its legal requirements, and 173 \emph{recitals} that provide 
additional context and clarifications to these articles. GDPR is an 
expansive set of regulation that covers the entire lifecycle of 
personal data. As such, achieving compliance requires interfacing with 
infrastructure components (including compute, network, and storage 
systems) as well as operational components (processes, policies, and 
personnel). However, since our investigation primarily concerns with 
GDPR's impact on storage systems, we focus on articles that describe 
the behavior of storage systems. These fall into two broad categories:
the rights of \emph{the data subjects} (i.e., the people whose 
personal data has been collected) and the responsibilities of \emph{
the data controllers} (i.e., the companies that collect 
personal data).

\subsection{Rights of the Data Subject} 
\label{sec-gdpr-rights}

There are 12 articles that codify the rights and freedoms of people. 
Among these, four directly concern storage systems.

The first one, \textsl {Article 15: \textsc{Right of access by the 
data subject}} allows any person whose personal data has been 
collected by a company to obtain detailed information about its usage 
including (i) the purposes of processing, (ii) the recipients to whom 
it has been disclosed, (iii) the period for which it will be stored, 
and (iv) its use in any automated decision-making. Thus, the storage 
system should not only be designed to store these metadata but also 
be organized to allow a timely access. Related to this is the 
\textsl {Article 21: \textsc{Right to object}}, which allows a person 
to object at any time to using their personal data for the purposes 
of marketing, scientific research, historical archiving, or profiling. 
This requires storage systems to know both whitelisted and blacklisted 
purposes associated with personal data at all times, and control 
access to it dynamically.

However, prominently, \textsl{Article 17: \textsc{Right 
to be forgotten}} grants people the right to require the data 
controller to erase their personal data without undue delay\footnote
{Article 17 covers only the personal data, not the insights 
derived from it; nor can it be used to violate the rights 
of other people or law enforcement.}. This right is broadly construed 
whether or not the personal data was obtained directly from the 
customer, or if the customer had previously given consent. From a
storage perspective, the article demands that the requested data be 
erased in a timely manner including all its replicas and backups. 
Finally, \textsl{Article 20: \textsc{Right to data portability}} 
states that people have the right to obtain all their personal 
information in a commonly used format as well as 
the right to have these transmitted to another company directly. Thus, 
storage systems should have the capability to access and transmit all 
data belonging to a particular user in a timely fashion. 


\begin{table*}[t]
\makebox[1\textwidth][c]{
\begin{minipage}[b]{1\textwidth}
\centering
\small
{\renewcommand{\arraystretch}{1.1}
\begin{tabular}{| c | l | l | l |} \hline
\thead{\bf No.} & \thead{\bf GDPR article} & \thead{\bf Key requirement} & \thead{\bf Storage feature} \\ \hline \hline
{5.1} & Purpose limitation & Data must be collected and used for specific purposes & Metadata indexing \\ \hline
{5.1} & Storage limitation & Data should not be stored beyond its purpose & Timely deletion \\ \hline
{5.2} & Accountability & Controller must be able to demonstrate compliance & All \\ \hline
{13} & Conditions for data collection & Get user's consent on how their data would be managed & All \\ \hline
{15} & Right of access by users & Provide users a timely access to all their data & Metadata indexing \\ \hline
{17} & Right to be forgotten & Find and delete groups of data & Timely deletion \\ \hline
{20} & Right to data portability & Transfer data to other controllers upon request & Metadata indexing \\ \hline
{21} & Right to object & Data should not be used for any objected reasons & Metadata indexing \\ \hline
{25} & Protection by design and by default & Safeguard and restrict access to data & Access control, Encryption \\ \hline
{30} & Records of processing activity & Store audit logs of all operations & Monitoring \\ \hline
{32} & Security of data & Implement appropriate data security measures & Access control, Encryption \\ \hline
{33, 34} & Notify data breaches & Share insights and audit trails from concerned systems & Monitoring \\ \hline 
{46} & Transfers subject to safeguards & Control where the data resides & Manage data location \\ \hline
\end{tabular}
}
\end{minipage}}
\caption{\emph {Key GDPR articles that significantly impact the 
design, interfacing, or performance of storage systems. The table 
maps the requirements of these articles into storage system features.}}
\vspace{-0.4cm}
\label{fig:regulation-table}
\end{table*}

\subsection{Responsibilities of the Data Controller} 
\label{sec-gdpr-responsibilities}

Among the articles that outline the responsibilities of data 
controllers, 10 concern storage systems. 

Three articles elucidate the high-level principles of data security 
and privacy that must be followed by all controllers. \textsl{Article 
24: \textsc{Responsibility of the controller}} establishes that the 
ultimate responsibility for the security of all personal data lies 
with the controller that has collected it; \textsl{Article 32: 
\textsc{Security of processing}} requires the controller to implement 
risk-appropriate and state-of-the-art security measures including 
encryption and pseudonymization; and lastly, \textsl{Article 25: 
\textsc{Data protection by design and by default}}, specifies that 
all systems must be designed, configured, and administered with data 
protection as a primary goal.

There are several articles that set guidelines for the collection, 
processing, and transmission of personal data. The purpose limitation 
of \textsl{Article 5: \textsc{Processing of personal data}} mandates 
that personal data should only be collected for specific purposes and 
not be used for any other purposes. From a storage point, this 
translates to maintaining associated (purpose-)metadata that could be 
accessed and updated by systems that process personal data. 
Interestingly, \textsl{Article 13} also ascertains that data subjects 
have the right to know the specific purposes for which their personal 
data would be used as well as the duration for which it will be stored. 
The latter requirement means that storage systems have to support 
time-to-live mechanisms in order to automatically erase the expired 
personal data.

Finally, while \textsl{Article 30: \textsc{Records of processing 
activities}} requires the controller to maintain logs of all 
activities concerning personal data, \textsl{Article 33: 
\textsc{Notification of personal data breach}} mandates them to notify 
the authorities and users within 72 hours of any personal data breaches. 
In conjunction with Accountability clause of \textsl{Article 5} which 
puts the onus of proving compliance on the controller, these articles 
impose stringent requirements on storage systems: to monitor and 
maintain detailed logs of all control- and data-paths interactions. 
For instance, every read operation now has to be followed by a 
(logging-)write operation. 

Table--\ref{fig:regulation-table} summarizes these articles and
translates their key requirements into specific storage features.

\section{Designing for Compliance}
\label{sec-design}

Based on our analysis of GDPR, we identify six key features that 
a storage system must support to be GDPR-compliant. Then,
we characterize how systems show variance in their support for 
these features.

\subsection{Features of GDPR-Compliant Storage}
\label{sec-gdpr-storage-features}

\vheading{Timely Deletion}. Under GDPR, no personal data can be retained 
for an indefinite period of time. Therefore, the storage system should 
support mechanisms to associate time-to-live (TTL) counters for 
personal data, and then automatically erase them from all internal 
subsystems in a timely manner. GDPR allows TTL to be either a static 
time or a policy criterion that can be objectively evaluated. 

\vheading{Monitoring and Logging}. In order to demonstrate compliance, 
the storage system needs an audit trail of both its internal actions 
and external interactions. Thus, in a strict sense, all operations 
whether in the data path (say, read or write) or control path (say, 
changes to metadata or access control) needs to be logged.
 
\vheading{Indexing via Metadata}. Storage systems should have 
interfaces to allow quick and efficient access to groups of data. For
example, accessing all personal data that could be processed under a 
specific purpose, or exporting all data belonging to a user. 
Additionally, it should have the ability to quickly 
retrieve and delete large amounts of data that match a criterion.

\vheading{Access Control}. As GDPR aims to limit access to personal 
data to only permitted entities, for established purposes, and for 
predefined duration of time, the storage system must support
fine-grained and dynamic access control.

\vheading{Encryption}. GDPR mandates that personal data be encrypted 
both at rest and in transit. While pseudonymization may help reduce 
the scope and size of data needing encryption, it is still required 
and likely results in degradation of storage system performance. 

\vheading{Managing Data Location}. Finally, GDPR restricts the 
geographical locations where personal data may be stored. This implies 
that storage systems should provide an ability to find and control the 
physical location of data at all times.

\subsection{Degree of Compliance}
\label{sec-design-degree}

Though GDPR is clear in its high-level goals, it is
intentionally vague in its technical specifications. For example, GDPR 
mandates that no personal data can be stored indefinitely and must be 
deleted after its expiry time. However, it does not specify how soon 
after its expiry should the data be erased? Seconds, hours, or even 
days? GDPR is silent on this, only mentioning that the data should be 
deleted without an undue delay. What this means for system designers 
is that GDPR compliance need not be a fixed target, instead a 
spectrum. We capture this variance along two dimensions: 
\emph{response time} and \emph{capability}.

\vheading{Real-time vs. Eventual Compliance}. Real-time compliance is
when a system completes the GDPR task (\eg deleting expired data or 
responding to user queries) synchronously in real-time. Otherwise, 
we categorize it as eventually compliant. Given the steep penalties 
(up to 4\% of global revenue or \euro20M, whichever is higher) for 
violating compliance, companies would do well to be in the strict end 
of the spectrum. However, as we demonstrate in \sref{sec-redis}, 
achieving real-time compliance results in significantly high overhead 
unless the challenges outlined in \sref{sec-discuss-research} are 
solved. This problem is further exacerbated for organizations that 
operate at scale. For example, Google cloud platform 
informs~\cite{google-cloud-deletes} their users that for a deleted 
data to be completely removed from all their internal systems, it 
could take up to 6 months.

\vheading{Full vs. Partial Compliance}. Distinct from the response 
time, systems exhibit varying levels of feature granularities 
and capabilities. Such discrepancies arise because many 
GDPR requirements sit at odds with the design principles and 
performance guarantees of certain systems. For example, file systems 
do not implement indexing into files as a core operation since that 
feature is commonly supported via application software like {\tt 
grep}. Similarly, many relational databases only partially and 
indirectly support TTL as that operation could be realized using 
user-defined triggers, albeit inefficiently. Thus, we define 
\emph{{full compliance}} to be natively supporting all the GDPR 
features, and \emph{{partial compliance}} as enabling feature support 
in conjunction with external infrastructure or policy components.

We use the term \emph{strict compliance} to reflect that a system
has achieved both full- and real-time compliance. 

\section{GDPR-Compliant Redis}
\label{sec-redis}

Redis~\cite{redis} is a prominent example of key-value stores, a
class of storage where unstructured data (i.e., value) is stored in 
an associative array and indexed by unique keys. Our choice of Redis 
as the reference system is motivated by two reasons: (i) it is
a modern storage system with an active open-source development, and 
(ii) key-value stores, in general, are not only widely deployed in 
Internet-scale systems ~\cite{dynamo-amazon, voldemort-linkedin, 
memcached-facebook} but are also an active area of research 
~\cite{hashcache-nsdi, skimpystash-sigmod, silt-sosp, 
hyperdex-sigcomm, lsm-trie-atc, triad-atc, pebbles-sosp}. 

From amongst the features outlined in \sref{sec-gdpr-storage-features},
Redis fully supports {monitoring}, {metadata indexing}, and 
{managing data locations}; partially supports {timely 
deletion}; offers no native support for {access control} and 
{encryption}. Below, we discuss our changes---some 
involving implementation while others simply concerning 
policy and configurations---towards making Redis, \emph{GDPR 
compliant}. This effort resulted in $\sim$120 lines of code
and configuration changes within Redis. 

Then, we evaluate the performance impact of our modifications to
Redis (v4.0.11) using the Yahoo Cloud Serving Benchmark (YCSB)
~\cite{ycsb}. We configure YCSB workloads to use 2M operations, and 
run them on a Dell Precision Tower 7810 with quad-core Intel Xeon 
2.8GHz processor, 16 GB RAM, and 1.2TB Intel 750 SSD.

\begin{figure}[t]
\centering
\includegraphics[width=0.5\textwidth]{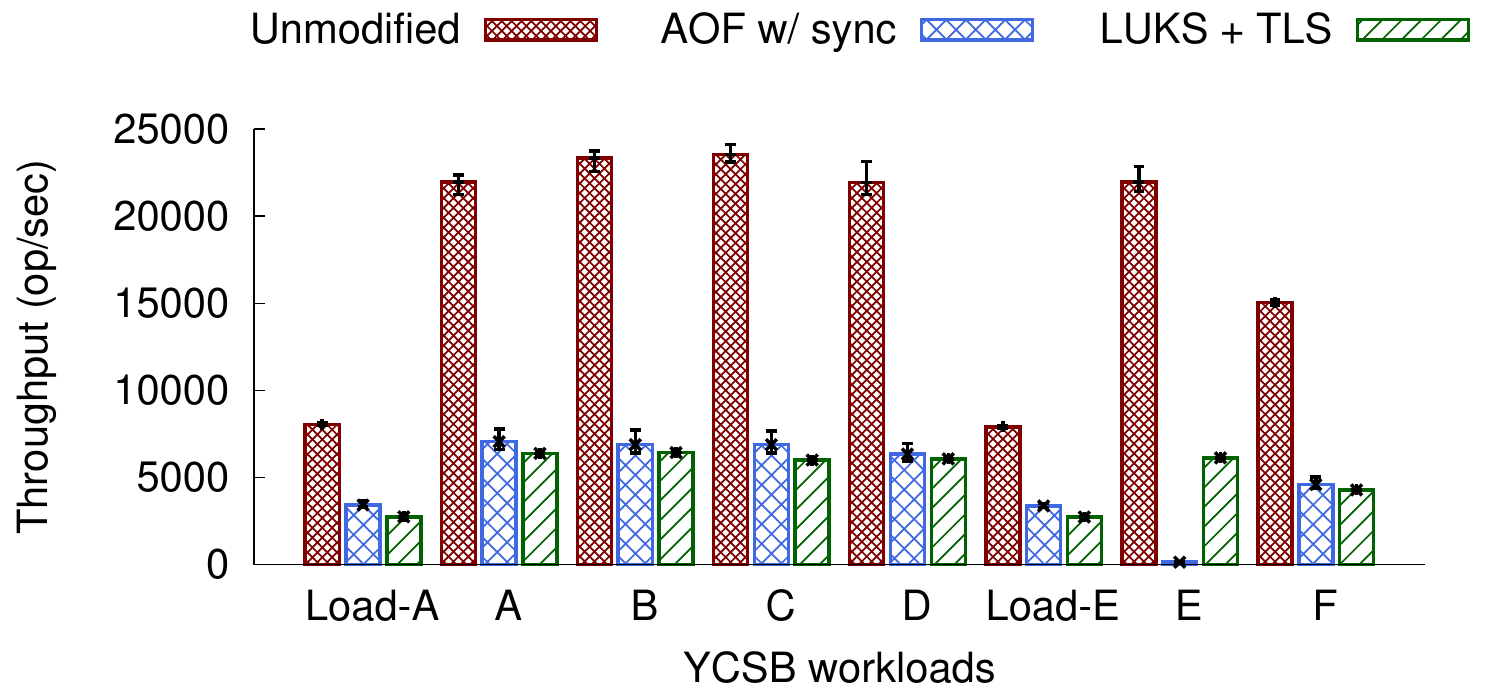}
\caption{\emph{Performance overhead of GDPR-compliant Redis. 
YCSB benchmarking shows that monitoring and encryption 
will each reduce Redis' throughput to $\sim$30\% of the original.}}
\vspace{-0.4cm}
\label{fig:redis-gdpr-overhead}
\end{figure}

\subsection{Monitoring and Logging}
\label{sec-redis-monitoring}

Redis offers several mechanisms to generate complete audit logs: a
debugging command called {\tt MONITOR}, configuring the server with 
slowlog option, and piggybacking on append-only-file ({\tt AOF}). Our 
microbenchmarking revealed that since Redis anyway performs its 
journaling via {\tt AOF}, the first two options result in more overhead 
than {\tt AOF}. Also, {\tt MONITOR} streams the logs over a network, thus 
requiring additional encryption. So, we selected the {\tt AOF} approach.
However, {\tt AOF} records only those operations that modify the dataset. 
Thus, we had to update the {\tt AOF} code to include all of Redis' 
interactions. Our benchmarking shows that when we set {\tt AOF} to 
fsync every operation to the disk synchronously, Redis' throughput
drops to $\sim$5\% of its original. But as 
Figure~\ref{fig:redis-gdpr-overhead} shows, when we relaxed the 
fsync frequency to once every second, the performance improved by 
6\myx i.e., throughput dropped only to $\sim$30\% the original.

\viheading{Key takeaway}: Even fully supported features like
\emph{logging} can cause significant performance overheads. 
Interestingly, the overheads vary significantly based on how 
strictly the compliance is enforced.

\subsection{Encryption}
\label{sec-redis-encryption}

In lieu of natively extending Redis' limited security model, we 
incorporate third-party modules for encryption. For data at rest, we 
use the Linux Unified Key Setup (LUKS)~\cite{luks}, and for data in 
transit, we set up transport layer security (TLS) using Stunnel
~\cite{stunnel}. Figure~\ref{fig:redis-gdpr-overhead} shows that 
Redis performs at a third of its original throughput when encryption 
is enabled. We observed that most of overhead was due to TLS: this 
was because the TLS proxies in our setup had reduced the average 
available network bandwidth from 44 Gbps to 4.9 Gbps, thereby 
affecting both latency and throughput of YCSB. While there are 
alternatives to the LUKS-TLS approach like key-level encryption, 
our investigation using the open-source Themis~\cite{themis} 
cryptographic library showed similar performance overheads.

\viheading{Key takeaway}: Retrofitting new features, especially those
that do not align with the core design philosophies, will result in 
excessive performance overheads.

\subsection{Timely Deletion}
\label{sec-redis-ttl}

While GDPR does not mandate a timeline for erasing the personal 
data after a request has been issued, it does specify that such data 
be removed from everywhere without undue delays. Redis offers three 
groups of primitives to erase data: (i) {\tt DEL} \& {\tt UNLINK} to 
remove one or more specified keys immediately, (ii) {\tt EXPIRE} \& 
{\tt EXPIREAT} to delete a given key after a specified timeout period, 
and (iii) {\tt FLUSHDB} \& {\tt FLUSHALL} to delete all the keys 
present in a given database or all existing databases respectively. 
The current mechanisms and policies of Redis present two hindrances. 

The first issue concerns the lag between the time of request and time 
of actual removal. While most of the above commands erase the data 
proactively, taking a time proportional to the size of data being 
removed, {\tt EXPIRE*} commands take a passive approach. The only way 
to guarantee the removal of an expired key is for a client to 
proactively access it. In absence of this, Redis runs a lazy 
probabilistic algorithm: once every 100ms, it samples 20 random keys 
from the set of keys with expire flag set; if any of these twenty have 
expired, they are actively deleted; if less than 5 keys got deleted, 
then wait till the next iteration, else repeat the loop immediately. 
Thus, as percentage of keys with associated expire increases, the 
probability of their timely deletion decreases.

\begin{figure}[t]
\centering
\includegraphics[width=0.48\textwidth]{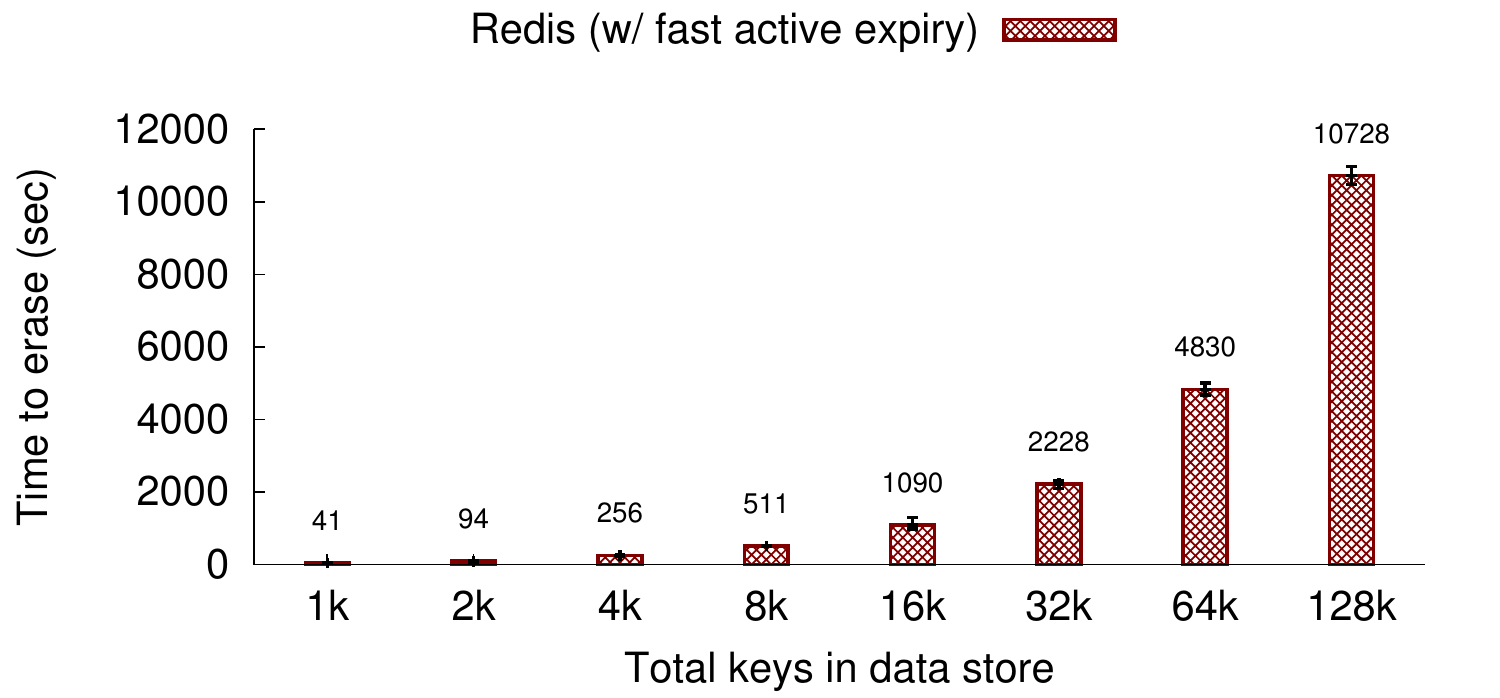}
\caption{\emph{The graph shows the delay in erasing the expired keys
(20\% of total keys in each case) beyond their TTL. In contrast, our 
GDPR-compliant Redis erases all the expired keys within sub-second 
latency.}}
\vspace{-0.4cm}
\label{fig:delay-expiry}
\end{figure}

To quantify this delay in erasure, we populate Redis with keys, all of
which have an associated expiry time. The time-to-live values are set
up such that 20\% of the keys will expire in short-term (5 minutes)
and 80\% in the long-term (5 days). Figure~\ref{fig:delay-expiry} then
shows the time Redis took to completely erase the short-term keys once
5 minutes have elapsed. As expected, the time to erasure increases
with the database size. For example, when there are 128k keys, clean up
of expired keys ($\sim$25k of them) took nearly 3 hours. To support a
stricter compliance, we modify Redis to iterate through the entire 
list of keys with associated {\tt EXPIRE}. Then, we re-run the same 
experiment to verify that all the expired keys are erased within 
sub-second latency for sizes of up to 1 million keys.

The second concern relates to the persistence of deleted data in 
subsystems beyond the main storage engine. For example, in Redis AOF 
persistence model, any deleted data persists in AOF until its 
compaction either via a policy-triggered or user-induced 
{\tt BGREWRITEAOF} operation. Though Redis prevents any legitimate 
access to data that is already deleted, its decision to let these 
persist in various subsystems, purely for performance reasons, is 
antithetical to the GDPR goals besides exposing itself to side-channel 
attacks. A naive approach to guaranteeing an immediate removal of 
deleted personal data is to trigger AOF compaction every time a key 
gets deleted. However, since GDPR only mandates a reasonable time for 
clean up, it may be prudent to configure a periodic (say, hourly) AOF 
compaction, which in turn would guarantee that no deleted key persists 
beyond an hour boundary.

\viheading{Key takeaway}: Even when the system supports a GDPR 
feature, system designers should carefully analyze its internal data 
structures, algorithms, and configuration parameters to gauge the 
degree of compliance.

\section{Concluding Remarks}
\label{sec-discussion}

We analyze the impact of GDPR on storage systems. We find that
achieving strict compliance efficiently is hard; a naive attempt at
strict compliance results in significant slowdown. We modify
Redis to be GDPR-compliant and measure the performance overhead of
each modification. Below, we identify three key research challenges 
that must be addressed to achieve strict GDPR compliance efficiently.

\subsection{Research Challenges}
\label{sec-discuss-research}

\vheading{Efficient Logging}. For strict compliance, every storage 
operation including reads must be synchronously written to persistent 
storage; persisting to solid state drives or hard drives results in 
significant performance degradation. New non-volatile memory 
technologies, such as Intel 3D Xpoint, can help reduce such overheads.
Efficient auditing may also be achieved through the use of eidetic 
systems. For example, Arnold~\cite{eidetic-systems} is able to 
remember past state with only 8\% overhead; adapting Arnold for GDPR 
remains a challenge.

\vheading{Efficient Deletion}. With all personal data possessing an 
expiry timestamp, we need data structures to efficiently find and 
delete (possibly large amounts of) data in a timely manner. Like 
timeseries databases, data can be indexed by their expiration time, 
then grouped and sorted by that index to speed up this process. 
However, GDPR is vague in its interpretation of deletions: it neither 
advocates a specific timeline for completing the deletions nor 
mandates any specific techniques. Thus, it remains to be seen if 
efforts like Google cloud's guarantee~\cite{google-cloud-deletes} to 
not retain customer data after 180 days of delete requests  
be considered compliant behavior.

\vheading{Efficient Metadata Indexing}. Several articles of GDPR
require efficient access to groups of data based on certain 
attributes. For example, accessing all the keys that allow processing
for a particular \emph{purpose} while ignoring those that object to 
that purpose; or collating all the files of a particular \emph{user} 
to be ported to a new controller. While traditional databases natively 
offer this ability via secondary indices, not all storage systems have
efficient or configurable support for this capability.

\subsection{Limitations and Importance}
\label{sec-discuss-generalize}

Given its preliminary nature, our work has several limitations. 
First, we investigate one particular storage system, Redis,
using one benchmark suite, YCSB. Expanding the scope to
a broader range of storage systems like relational databases
and file systems would increase the confidence of our findings.
Next, it is likely that the performance of our GDPR-compliant 
Redis could be further improved with a deeper knowledge of Redis 
internals. Finally, while we focus exclusively on storage systems, 
researchers have shown~\cite{gdpr-sins} how GDPR compliance 
requires organization wide changes to the systems that process
personal data. 

With the growing relevance of privacy regulations around the 
world, we expect this paper to trigger interesting conversations.
This is one of the first efforts to systematically
analyze the impact of GDPR on storage systems. We would be keen to 
engage the storage community in identifying and addressing the 
research challenges in this space.

{
\small
\bibliographystyle{plain}
\bibliography{paper}

\begin{thebibliography}{10}

\bibitem{luks}
Cryptsetup and {LUKS} - open-source disk encryption.
\newblock \url{https://gitlab.com/cryptsetup/cryptsetup}, Accessed May 2019.

\bibitem{google-cloud-deletes}
{Data Deletion on Google Cloud Platform}.
\newblock \url{https://cloud.google.com/security/deletion/}, Accessed May 2019.

\bibitem{redis}
Redis {D}ata {S}tore.
\newblock \url{https://redis.io}, Accessed May 2019.

\bibitem{stunnel}
Stunnel.
\newblock \url{https://www.stunnel.org}, Accessed May 2019.

\bibitem{themis}
Themis.
\newblock \url{https://github.com/cossacklabs/themis}, Accessed May 2019.

\bibitem{hashcache-nsdi}
Anirudh Badam, KyoungSoo Park, Vivek Pai, and Larry Peterson.
\newblock {HashCache: Cache Storage for the Next Billion.}
\newblock In {\em USENIX NSDI}, 2009.

\bibitem{triad-atc}
Oana Balmau, Diego Didona, Rachid Guerraoui, Willy Zwaenepoel, Huapeng Yuan,
  Aashray Arora, Karan Gupta, and Pavan Konka.
\newblock {TRIAD: Creating synergies between memory, disk and log in log
  structured key-value stores}.
\newblock In {\em USENIX ATC}, 2017.

\bibitem{ycsb}
Brian Cooper, Adam Silberstein, Erwin Tam, Raghu Ramakrishnan, and Russell
  Sears.
\newblock {Benchmarking cloud serving systems with YCSB}.
\newblock In {\em ACM SoCC}, 2010.

\bibitem{skimpystash-sigmod}
Biplob Debnath, Sudipta Sengupta, and Jin Li.
\newblock {SkimpyStash: RAM space skimpy key-value store on flash-based
  storage}.
\newblock In {\em ACM SIGMOD}, 2011.

\bibitem{dynamo-amazon}
Giuseppe DeCandia, Deniz Hastorun, Madan Jampani, Gunavardhan Kakulapati,
  Avinash Lakshman, Alex Pilchin, Swaminathan Sivasubramanian, Peter Vosshall,
  and Werner Vogels.
\newblock Dynamo: {A}mazon's {H}ighly {A}vailable {K}ey-{V}alue {S}tore.
\newblock In {\em USENIX OSDI}, 2007.

\bibitem{eidetic-systems}
David Devecsery, Michael Chow, Xianzheng Dou, Jason Flinn, and Peter~M Chen.
\newblock {Eidetic Systems}.
\newblock In {\em USENIX OSDI}, 2014.

\bibitem{hyperdex-sigcomm}
Robert Escriva, Bernard Wong, and Emin~G{\"u}n Sirer.
\newblock {HyperDex: A distributed, searchable key-value store}.
\newblock In {\em ACM SIGCOMM}, 2012.

\bibitem{gartner-prediction}
Amy~Ann Forni and Rob van~der Meulen.
\newblock Organizations are unprepared for the 2018 {E}uropean {D}ata
  {P}rotection {R}egulation. {I}n \emph{Gartner}, May 2017.

\bibitem{equifax}
Todd Haselton.
\newblock Credit reporting firm equifax says data breach could potentially
  affect 143 million {US} consumers. {I}n \emph{CNBC}, Sep 7 2017.

\bibitem{silt-sosp}
Hyeontaek Lim, Bin Fan, David Andersen, and Michael Kaminsky.
\newblock {SILT: A} memory-efficient, high-performance key-value store.
\newblock In {\em ACM SOSP}, 2011.

\bibitem{memcached-facebook}
Rajesh Nishtala, Hans Fugal, Steven Grimm, Marc Kwiatkowski, Herman Lee, Harry
  Li, Ryan McElroy, Mike Paleczny, Daniel Peek, Paul Saab, David Stafford, Tony
  Tung, and Venkateshwaran Venkataramani.
\newblock Scaling {M}emcache at {F}acebook.
\newblock In {\em USENIX NSDI}, 2013.

\bibitem{pebbles-sosp}
Pandian Raju, Rohan Kadekodi, Vijay Chidambaram, and Ittai Abraham.
\newblock Pebbles{DB}: {B}uilding {K}ey-{V}alue {S}tores using {F}ragmented
  {L}og-{S}tructured {M}erge {T}rees.
\newblock In {\em ACM SOSP}, 2017.

\bibitem{gdpr-regulation}
General Data~Protection Regulation.
\newblock Regulation ({EU}) 2016/679 of the {E}uropean {P}arliament and of the
  {C}ouncil of 27 {A}pril 2016 on the protection of natural persons with regard
  to the processing of personal data and on the free movement of such data, and
  repealing {D}irective 95/46.
\newblock {\em Official Journal of the European Union}, 59(1-88), 2016.

\bibitem{data-breaches-2017}
Victor Reklaitis.
\newblock How the number of data breaches is soaring. {I}n \emph{MarketWatch},
  May 25 2018.

\bibitem{gdpr-sins}
Supreeth Shastri, Melissa Wasserman, and Vijay Chidambaram.
\newblock {The Seven Sins of Personal-Data Processing Systems under GDPR}.
\newblock In {\em USENIX HotCloud}, 2019.

\bibitem{cambridge-analytica}
Olivia Solon.
\newblock {Facebook says Cambridge Analytica may have gained 37{M} more users'
  data}. {I}n \emph{The Guardian}, Apr 4 2018.

\bibitem{voldemort-linkedin}
Roshan Sumbaly, Jay Kreps, Lei Gao, Alex Feinberg, Chinmay Soman, and Sam Shah.
\newblock Serving {L}arge-scale {B}atch {C}omputed {D}ata with {P}roject
  {V}oldemort.
\newblock In {\em USENIX FAST}, 2012.

\bibitem{lsm-trie-atc}
Xingbo Wu, Yuehai Xu, Zili Shao, and Song Jiang.
\newblock {LSM-trie: an LSM-tree-based Ultra-Large Key-Value Store for Small
  Data}.
\newblock In {\em USENIX ATC}, 2015.

\end{thebibliography}
}

\end{document}